\documentclass[preprint]{vgtc}              

\onlineid{1508}

\vgtccategory{Research}

\vgtcpapertype{application/design study}

\title{"They Aren't Built For Me": An Exploratory Study of Strategies for Measurement of Graphical Primitives in Tactile Graphics}

\author{\authororcid{Areen Khalaila}{0009-0006-8166-9794}\thanks{e-mail: areenkh@brandeis.edu}\\ %
        \scriptsize Brandeis University %
\and \authororcid{Lane Harrison}{0000-0003-3029-2799}\thanks{e-mail: ltharrison@wpi.edu}\\ %
     \scriptsize Worcester Polytechnic Institute %
\and \authororcid{Nam Wook Kim}{0000-0003-4899-6671}\thanks{e-mail: nam.wook.kim@bc.edu}\\ %
     \scriptsize Boston College %
\and \authororcid{Dylan Cashman}{0000-0003-4853-5701}\thanks{e-mail: dylancashman@brandeis.edu}\\ %
     \parbox{1.4in}{\scriptsize \centering Brandeis University}}

\abstract{%
Advancements in accessibility technologies such as low-cost swell form printers or refreshable tactile displays promise to allow blind or low-vision (BLV) people to analyze data by transforming visual representations directly to tactile representations.
However, it is possible that design guidelines derived from experiments on the visual perception system may not be suited for the tactile perception system.
We investigate the potential mismatch between familiar visual encodings and tactile perception in an exploratory study into the strategies employed by BLV people to measure common graphical primitives converted to tactile representations.
First, we replicate the Cleveland and McGill study on graphical perception using swell form printing with eleven BLV subjects.  Then, we present results from a group interview in which we describe the strategies used by our subjects to read four common chart types.  
While our results suggest that familiar encodings based on visual perception studies can be useful in tactile graphics, our subjects also expressed a desire to use encodings designed explicitly for BLV people.  Based on this study, we identify gaps between the perceptual expectations of common charts and the perceptual tools available in tactile perception.  Then, we present a set of guidelines for the design of tactile graphics that accounts for these gaps.
Supplemental material is available at \url{https://osf.io/3nsfp/?view_only=7b7b8dcbae1d4c9a8bb4325053d13d9f}.
}


\teaser{
  \centering
  \includegraphics[width=\linewidth, alt={Metaphorical icons representing five strategies outlined in the text: In tactile perception, translatable strategies exist for some types of measurement such as length and angle but not all measurements, such as area and popout.}]{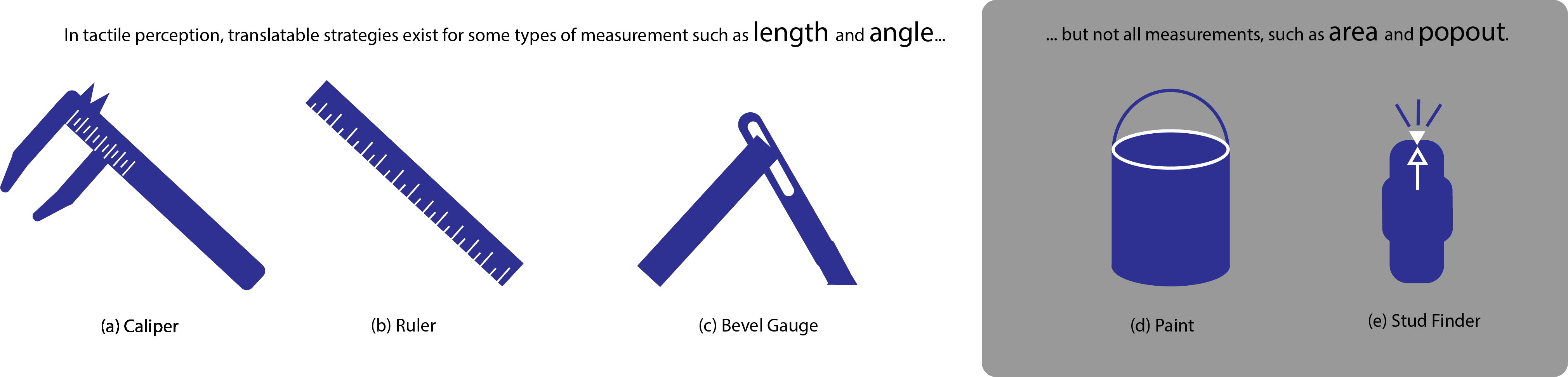}
  \caption{%
  	Five strategies utilized within our study for measurements of tactile stimuli in a replication of the Cleveland and McGill experiment.  In each strategy, hands and fingers are used to measure the physical channel.  Strategies (a), (b), and (c) can aptly measure length and angle, but strategies (d) (used to measure area) and (e) (used to recognize popout) are mismatched, implying additional encoding rules in tactile representations.%
  }
  \label{fig:teaser}
}

\graphicspath{{tactile_graphics_strategies/figs/}{./}} 

\usepackage{tabu}                      
\usepackage{booktabs}                  
\usepackage{lipsum}                    
\usepackage{mwe}                       

\usepackage{mathptmx}                  
\usepackage{amsmath}
\usepackage{xspace}
\usepackage{wrapfig}
\usepackage{titlesec}
\usepackage{dirtytalk}

\titlespacing\section{0pt}{12pt plus 1pt minus 1pt}{0pt plus 2pt minus 2pt}
\titlespacing\subsection{0pt}{12pt plus 1pt minus 1pt}{0pt plus 2pt minus 2pt}

\begin{document}


\firstsection{Introduction}

\maketitle

There is a growing requirement within daily life to engage with graphical representations of data.  A recent example is the \textit{Flattening the Curve} campaign during the COVID-19 pandemic, which relied on graphical representations to convey the urgency of social distancing measures \cite{thunstrom2020benefits}. This explicitly visual message excluded blind or low vision (BLV) people from fully engaging with data-centric discourse that was highly relevant to their lives. The adaptation of such graphics into non-visual modalities often occurs as an afterthought, leading to disparities in information access. Although alternative representations, such as sonification and tactile graphics, have been explored, there remains a significant gap in understanding how perceptual differences impact the effectiveness of these adaptations.

The challenge of making data visualization accessible extends beyond communication—it also affects economic and professional opportunities. As fields such as data science and analytics become increasingly integral to the job market, exclusion from standard visualization techniques can contribute to employment disparities. Studies show that BLV people face unemployment rates two to three times higher than the general population \cite{mcdonnall2019employment}. While initiatives such as the NSF’s Data Science Corps\footnote{\url{https://new.nsf.gov/funding/opportunities/data-science-corps-dsc}} and DARPA’s workforce training programs\footnote{\url{https://tools-competition.org/workforce/}} highlight the growing demand for data literacy, the accessibility of data visualization tools remains limited. Emerging consumer technologies, including refreshable tactile displays, attempt to bridge this gap by converting graphical information into non-visual formats. However, such adaptations often rely on conventions established for sighted users, without empirical validation of their effectiveness in tactile perception \cite{lighthouseFlattenCurve2021}.

A critical question remains: Should a tactile representation replicate its visual counterpart to be effective? Prior research suggests that while tactile and visual perception share some cognitive mechanisms, they also exhibit fundamental differences
related to the differing processing systems
\cite{Tabrik2022, Ault2002}. Yet many existing tactile visualizations are directly derived from sighted design principles without accounting for these differences, raising concerns about their usability for BLV individuals. While some studies have investigated tactile adaptations of visualizations \cite{pineros2025tactile, xu2023let}, this is the first study to our knowledge that investigates basic graphical perception experiments with BLV participants. Understanding the strategies employed by BLV users is crucial for designing more effective tactile representations that go beyond visual mimicry.

In this work, we investigate the strategies used by BLV participants in reading data from graphical primitives rendered in tactile media.  This study replicates the foundational work of Cleveland and McGill \cite{cleveland1984graphical} by exploring how graphical perception principles translate to tactile graphics for BLV users. Using swell paper, a widely used medium for tactile graphics in educational settings, we analyze how BLV participants interpret bar charts, pie charts, bubble charts, and stacked bar charts. Rather than focusing solely on accuracy and inference time comparisons with sighted users, we report the strategies BLV individuals employ to navigate and interpret these tactile encodings. 

Through a controlled study involving eleven BLV participants and a follow-up group interview, we identify a range of strategies—such as the use of finger \textit{calipers} for measuring bar lengths and \textit{rulers} for comparing distances—that shape how BLV users engage with tactile data. Our findings reveal that while some visual encoding principles translate well to tactile formats, others introduce significant perceptual challenges due to physical differences between perception with hands vs. eyes.
These challenges highlight gaps between design guidelines for visual primitives and the needs of BLV users in tactile media.  These gaps lead us to a set of guidelines for the design of effective visualizations for tactile graphics.

\section{Related Work}

\subsection{Graphical Perception in Data Visualization}

The graphical perception study from Cleveland and McGill~\cite{cleveland1984graphical} has significantly influenced the design and use of statistical charts and data visualizations. 
As the medium for viewing graphics has evolved and the ubiquity of visualizations has increased, researchers have replicated the study to observe whether differences appear in the performance of different groups of participants.
Heer and Bostock replicated Cleveland and McGill's graphical perception study using Amazon's Mechanical Turk, extending the original controlled laboratory experiments to a broader, more diverse participant pool.  
They reported similar performance to the original study, concluding that crowdsourced studies should produce comparable insights to laboratory studies. 
Other studies have extended graphical perception research to other non-traditional populations. 
For instance, Le et al.~\cite{le2014elementary} conducted an experiment comparing the graphical perception of a general population (under 60 years of age) with that of older adults (60 years and above). 
Panavas et al.~\cite{panavas2022juvenile} examined how children aged 8 to 12 perceive graphs compared to adult students, and Wu et al.~\cite{wu2021understanding} investigate how individuals with intellectual and developmental disabilities interpret different visualization designs.

For BLV individuals, particularly those who are completely blind, graphical perception experiments are not feasible due to the lack of visual input. 
As a result, alternative information modalities have been explored to assess their chart-reading capabilities; Kim et al. provide an overview of past studies on visualization accessibility for BLV individuals~\cite{kim2021accessible}. 
For instance, several studies have investigated how BLV users perceive auditory graphs. 
These include research on optimal polarity and scaling parameters in sonified data~\cite{walker2010universal}, comparisons between non-speech audio and speech-based representations~\cite{sakhardande2019comparing}, and examinations of different auditory channels (e.g., pitch, volume).

Several studies focus on the usability of tactile graphics—such as comparing different tactile representations~\cite{Goncu2010,Watanabe2018,Yalong2020}, evaluating tactile versus auditory modalities~\cite{Goncu2010}, and investigating the role of textures in tactile perception~\cite{Watanabe2018}—as well as haptic feedback approaches~\cite{paneels2009review,Danyang2022}. 
Screenreader software can serve as a medium for complementing or even replacing tactile graphics. 
Recent studies have explored how different formats impact data perception for screen readers, including data tables and keyboard navigation~\cite{kim2023beyond,zong2022rich}, conversational interactions~\cite{sharif2023understanding,kim2023beyond}, and audio narratives~\cite{Alexa2022}. 
Prakash et al.~\cite{Prakash2024} explored the graphical perception of bar charts by low-vision users, revealing that screen magnification poses unique challenges, such as increased perceptual effort and distortion effects.

While past studies have evaluated how BLV individuals interact with charts using alternative modalities, no direct comparison has been made between BLV and sighted individuals on graphical perception tasks.  In our work, we explore a direct comparison to better understand how assumptions baked into visual design might not be applicable to tactile graphics.  By replicating Cleveland and McGill in particular, which has been seminal in establishing foundational guidelines for effective visualization design, we hope to reveal new insights. 

\subsection{Studies in Tactile Graphics}

Tactile graphics enable BLV individuals to interact with graphical data through raised, textured surfaces. 
These non-visual representations translate visual elements like charts, maps, and diagrams into tactile formats using embossing, swell paper, or refreshable displays.

From the analysis of tactile charts from various resources, Engel and Weber~\cite{Engel2017} derived basic design guidelines for better readability, such as emphasizing the necessity of structured elements like tick marks, grid lines, and textured fill patterns. 
In a follow-up remote user study in which participants were asked to answer task questions such as finding extrema or comparing values, Engel and Weber~\cite{engel2019user} found no significant difference between birth-blind and late-blind participants while discussing design recommendations for standard charts such as bar, line, scatter, and pie charts. 
Watanabe and Inaba \cite{Watanabe2018} further deepened the understanding of tactile discriminability, identifying better texture granularity for tactile bar charts on capsule paper through experiments where participants were asked to count the number of bars under different texture conditions.

Other studies conducted comparative studies evaluating the effectiveness of different tactile modalities. Goncu et al.~\cite{Goncu2010} highlighted preferences for tactile diagrams with gridlines and Braille values over direct transcriptions, and tactile tables over tactile charts. 
Watanabe and Mizukami \cite{WatanabeMizukami2018} compare tactile scatter plots to tactile and electronic tables, where participants were asked to identify the relationship between two variables--- results showed that tactile graphs outperformed the other two conditions. 
In a related study with 8 BLV participants, Yang et al.~\cite{Yalong2020} compared tactile node-link and matrix diagrams for tasks such as pathfinding and cluster identification, noting a preference for node-link diagrams, which performed best except in adjacency tasks.

Our study contributes directly to this space by both bridging a gap between graphical perception studies by adapting the widely-replicated comparison task from Cleveland and McGill, and by extending prior studies by identifying unique “tool”-like approaches that participants use to answer low-level visualization tasks, which may be applicable to other visualizations and other visualization studies in the tactile graphics space. Our observed strategies mirror those identified by Perkins and Gardiner~\cite{Perkins01122003}, where participants engaged in two-handed exploration and edge-based scanning. These approaches suggest that spatial layout plays a critical role in tactile chart readability.


\subsection{Technologies for Tactile Graphics}

Traditional methods for creating tactile charts involve producing physical raised-line graphics that BLV readers can feel. Swell papers and Braille printers are commonly used to produce tactile charts~\cite{engel2019user}. Special embossing devices in Braille printers punch or raise dots on thick paper to form tactile lines and shapes, while swell paper machines produce the chart in black ink on capsule-coated paper and runs it through a fuser machine that heats the paper, causing the black areas to swell up into raised lines and textures.

Beyond these conventional methods, recent research has explored innovative approaches to tactile chart accessibility. Refreshable tactile displays (RTDs), such as Graphiti and Dot Pad, are dynamic pin-board or multi-line Braille displays that can raise and lower pins to render different graphics in real time.  MAIDR~\cite{seo2024maidr} introduces the concept of using a conventional 1D Braille device to create 2D charts, expanding the possibilities of tactile data representation. TactualPlot~\cite{chundury2023tactualplot} is an interaction technique that spatializes data as audio on a touchscreen; as the user's finger moves across different regions, the system generates corresponding sounds—varying in pitch, stereo pan, or timbre—to indicate the presence and value of data points. Similarly, ChartA11y~\cite{zhang2024charta11y} enhances chart accessibility on mobile devices by developing a multi-modal touch interface, allowing blind users to explore charts by touch while receiving audio feedback and vibration cues. Additionally, other audio-tactile interfaces have been proposed, including Slide-Tone and Tilt-Tone~\cite{fan2022slide}, Tactile Graphics Helper~\cite{fusco2015tactile}, GraVVITAs~\cite{goncu2011gravvitas}, and AudioFunctions~\cite{ahmetovic2019audiofunctions}, each contributing unique methods for making data visualizations more accessible to BLV individuals.  While RTDs present an exciting new medium for tactile graphics, they are only able to draw over a fixed course grid of pixels.  Findings on RTDs may not be applicable to swell form printing. In addition, they also are more expensive than swell form printing.  

Recently, Chen et al.~\cite{chen2025tactile} introduced a set of design guidelines for translating Vega-Lite charts into tactile formats based on interviews with tactile graphics designers.  Their work focuses on how to adapt existing visual encodings for non-visual exploration. Their guidelines resulted in design recommendations that could inform grammar-based tools like Tactile Vega-Lite.  While our work shares a similar motivation of the derivation of design guidelines in extending graphical perception principles to tactile contexts, we focus on design guidelines in general, rather than guidelines applicable to grammar-based tools. 
 In addition, we use different sources to develop our guidelines than that study. 
Our work specifically adds to this line of research by identifying strategies that people use when making judgments with tactile charts.

Closest to our present study, Xu et al.~\cite{xu2023let} conduct a within-subjects experiment comparing tactile and visual perceptual accuracy using swell paper charts. Their work compared the error rates of sighted participants on a graphical comparison task with visual representations and tactile representations.  
Notably, their study used sighted-but-blindfolded participants, while our study includes only BLV participants who have extensive familiarity and experience with tactile perception. 
This distinction enables us to explore how tactile perception operates in practice, particularly among individuals who routinely rely on touch to interpret information.




\section{Methodology}

The goal of this research is to observe the strategies used by BLV participants in making perceptual judgments of popular graphical primitives used in visualization design.  To accomplish this, we first chose a medium for tactile graphics.  Then, we chose which stimuli to evaluate, based on previous studies of graphical perception.

\noindent \textbf{Creation of Tactile Graphics:} 
The primary techniques for producing tactile graphics include embossing with a braille embosser, printing on swell paper, and thermoforming. Swell paper, utilized in our experiment, features microcapsules that expand when heated, creating raised features up to 0.5mm high \cite{Jonathan2003}. This method was chosen due to its precision, quick production capabilities, and cost-effective creation. The cost per tactile graphic was approximately \$2.75, making this approach economically viable for large-scale studies. Existing research based on an international survey of thirty blind and partially sighted people found a strong preference for swell paper, citing its enhanced tactile feedback and durability, making it ideal for users with vision impairments \cite{rowell2005feeling}. 


\begin{figure}[h]
  \centering
  \includegraphics[width=0.35\textwidth]{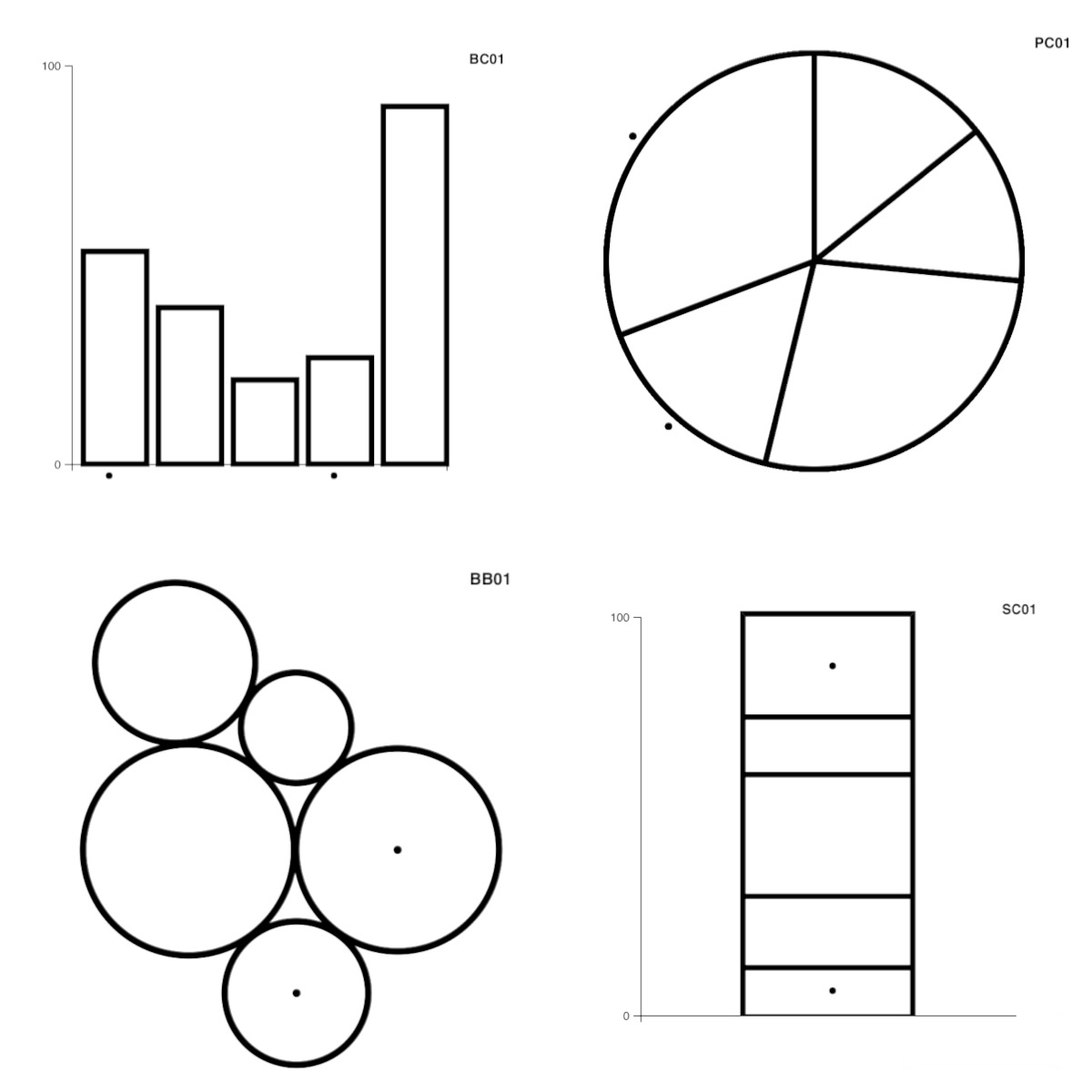}
  \caption{Four chart types used in our study as designed for swell form printing, each depicting a 50\% proportional relationship between two data elements: bar chart (top left), pie chart (top right), bubble chart (bottom left), and stacked bar chart (bottom right). Participants were asked to identify the two marked regions (indicated by central dots) and estimate what percentage the smaller was of the larger. 
  }
  \label{fig:tactile_stimuli}
\end{figure}



\noindent \textbf{Generating Stimuli:} 
We adapted the stimuli generation process from the original Cleveland and McGill study to accommodate tactile perception. 
In our study, we selected four common chart types: bar charts, pie charts, bubble charts, and stacked bar charts. These charts were chosen because they represent a mix of commonly used visual encodings—length, angle, area, and stacked length—and have been widely studied in graphical perception literature. Figure \ref{fig:tactile_stimuli} illustrates these chart types as rendered on swell paper.   

Some modifications were made to both the data generation method of Cleveland and McGill and the visual design of the primitives to accommodate this tactile format.  This involved constraining the ratios of the smaller to the larger graphical elements, which we refer to as the \textit{True Proportional Difference}, uniformly in 5\% increments from 50\% to 95\%, while allowing the larger elements to vary from 40\% to 100\% of the viewport. Distractor stimuli were also included, ranging from 20\% to 100\% of the viewport, to mimic real-world scenarios where multiple data points are presented together.  Compared to previous perception studies, we bound the minimum graphical size because in tactile graphics, there is typically lower resolution available, and lines need to be thick enough to print.  Because of this, very small graphics may not render correctly and were avoided in our study.  
In total, 80 different charts were produced, with two variations for each of the ten different values of True Proportional Difference (50\%, 55\%, \ldots, 90\%, 95\%), and four chart types.



\section{Experimental Design}



To understand how BLV users interpret tactile graphics, we conducted two studies. First, we replicated Cleveland and McGill’s foundational graphical perception study using tactile charts, examining whether patterns from visual studies hold in tactile contexts. Since many design decisions rely on their findings, any tactile differences could suggest a need for new design principles. We also collected qualitative data through think-aloud protocols. Two months later, we held a follow-up group interview where participants discussed strategies, challenges, and their overall experience, revealing design gaps not captured in the initial study.

\subsection{Replication Study: Measuring Accuracy and Response Time}
\textbf{Participant Recruitment and Setup:} 
In the first study, eleven participants in total were recruited from the community affiliated with a local school for the blind and the experiments were conducted over two days in a controlled environment. The participants ranged in age from 18 to over 65, with an average age of 48.3 years. The gender distribution included 6 males and 5 females. Education levels varied, with 5 participants having completed an undergraduate degree and 5 holding a master's degree, while 1 had some graduate education. Participants reported varying degrees of familiarity with tactile graphics. On a 7-point Likert scale (1 = Beginner, 7 = Expert), self-rated expertise with tactile displays ranged from 3 to 7 (\textit{M} = 5.91, \textit{SD} = 1.16). Most participants reported frequent to very frequent use of tactile graphics in their daily lives. Nine out of eleven participants had prior experience participating in studies related to accessibility or tactile perception, and six had received formal training in interpreting graphical data in tactile formats. Regarding vision status, participants described a range from total blindness to moderate or severe vision loss. Some noted residual light perception or shape discrimination under specific conditions. This variability may influence tactile exploration strategies and should be considered in interpreting performance. Each session lasted approximately 75 minutes and was conducted in person to enable effective interaction with the tactile graphics. 

\noindent \textbf{Procedure:} The procedure began with an informed consent briefing read aloud to participants, followed by a practice session with eight stimuli (two of each chart type) to familiarize them with the tactile graphics. Participants were then presented with the remaining 72 graphics in a randomized order (consistent across participants) to minimize physical degradation from repeated handling and asked to interpret these at their natural pace, verbalizing their thoughts and answers for audio recording.  Participants were told the true answer (percent size of the smaller stimuli to the larger) for their 8 training stimuli, but were not told during the experimental phase.  Answers were recorded into a spreadsheet during test time and then later adjudicated against recordings, with any disagreements removed from the dataset.  This phase of the experiment was limited to 30 minutes, and participants were notified when half of the time was left.

\noindent \textbf{Data Collection:}  Data were collected through both quantitative and qualitative methods. Quantitatively, we measured the accuracy of each response and the time taken to complete each task. Qualitatively, participants were asked to provide feedback about their experience, including any difficulties they encountered and their subjective assessment of the tactile graphs’ clarity and usability.  After the tactile interpretation tasks, participants completed a demographics survey and engaged in a 10-15 minute semi-structured interview. These interviews aimed to gather qualitative feedback on their experience and any challenges they faced during the experiment.  \footnote{Study documents, including interview questions, demographic questions, and graphics files are available at \url{https://osf.io/3nsfp/?view_only=7b7b8dcbae1d4c9a8bb4325053d13d9f}.}

\subsection{Follow-up Group Interview: Exploring Strategies and Challenges} Two months after the first study was conducted, the original participants were contacted via email to participate in a followup group interview in order to compare strategies for reading the tactile graphics.  Four participants took part in a second study held in a conference room for a 90 minute group interview.  For each of the four chart types, the participants were first each given two examples of the chart type (one close to a ground truth of 50\%, one close to a ground truth of 95\%) in order to refresh their memory about the chart type.  Then, they were each asked about their strategies for reading from the chart, what they found difficult about reading the chart, and any improvements they suggested.  The group interview ended by asking the four participants to respond to any of the strategies of the other participants for any of the chart types.  The meeting was photographed and its audio was recorded and transcribed for analysis of common strategies.


\section{Results}
\label{sec:supplement_inst}

\begin{figure}[h]
 \centering 
 \includegraphics[width=\columnwidth]{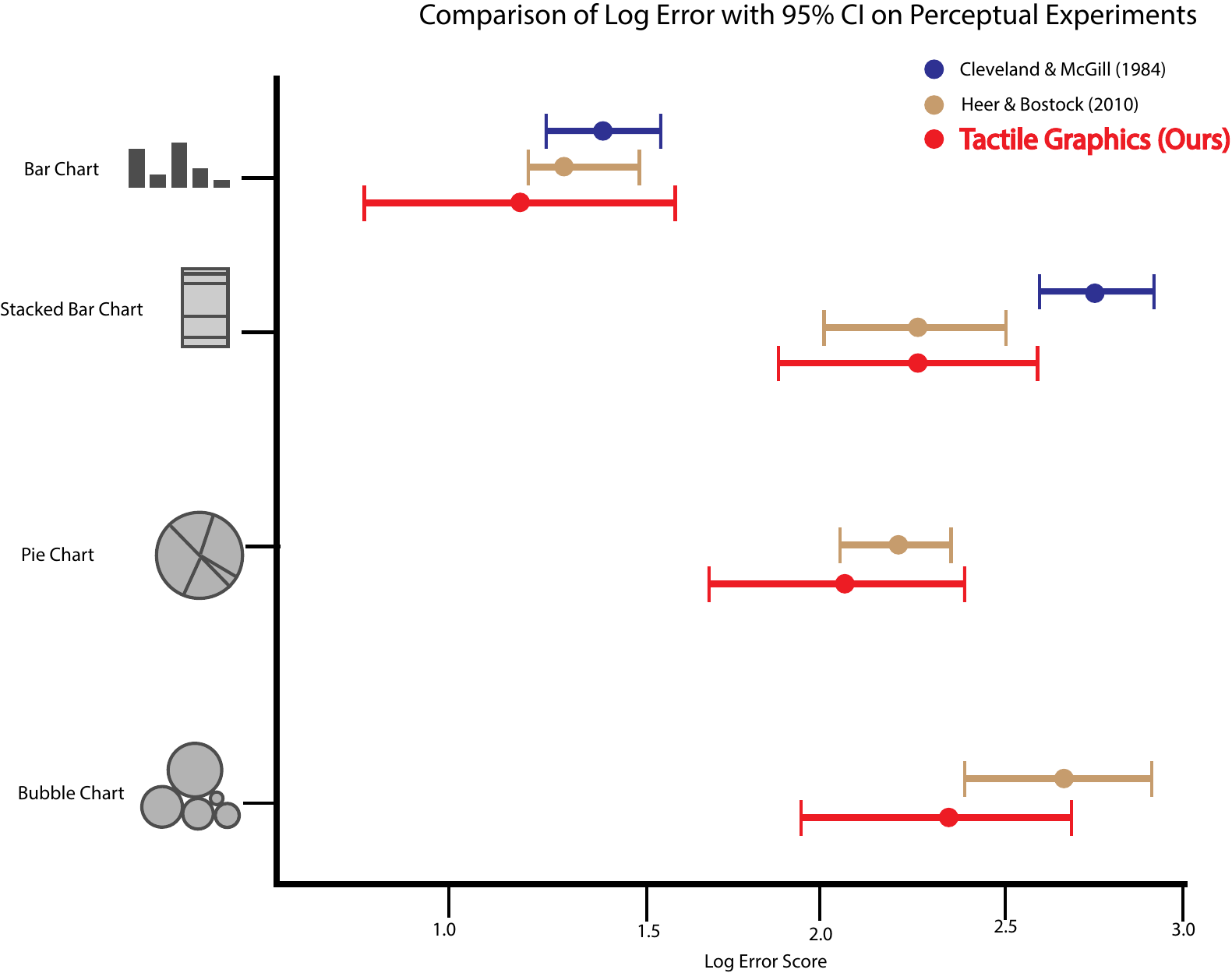}
 \caption{Error scores for our study (red) compared with two previous comparable studies: Cleveland and McGill~\cite{cleveland1984graphical} and Heer and Bostock~\cite{heer2010crowdsourcing}. Error bars indicate 95\% confidence intervals via bootstrapping.  Pie Chart and Bubble Chart were not measured in Cleveland and McGill.  Previous studies' results were derived via published graphics rather than raw data.}
 \label{fig:error_rates}
\end{figure}

\subsection{Error Scores by Chart Type}
\begin{figure}[h]
 \centering 
 \includegraphics[width=0.8\columnwidth]{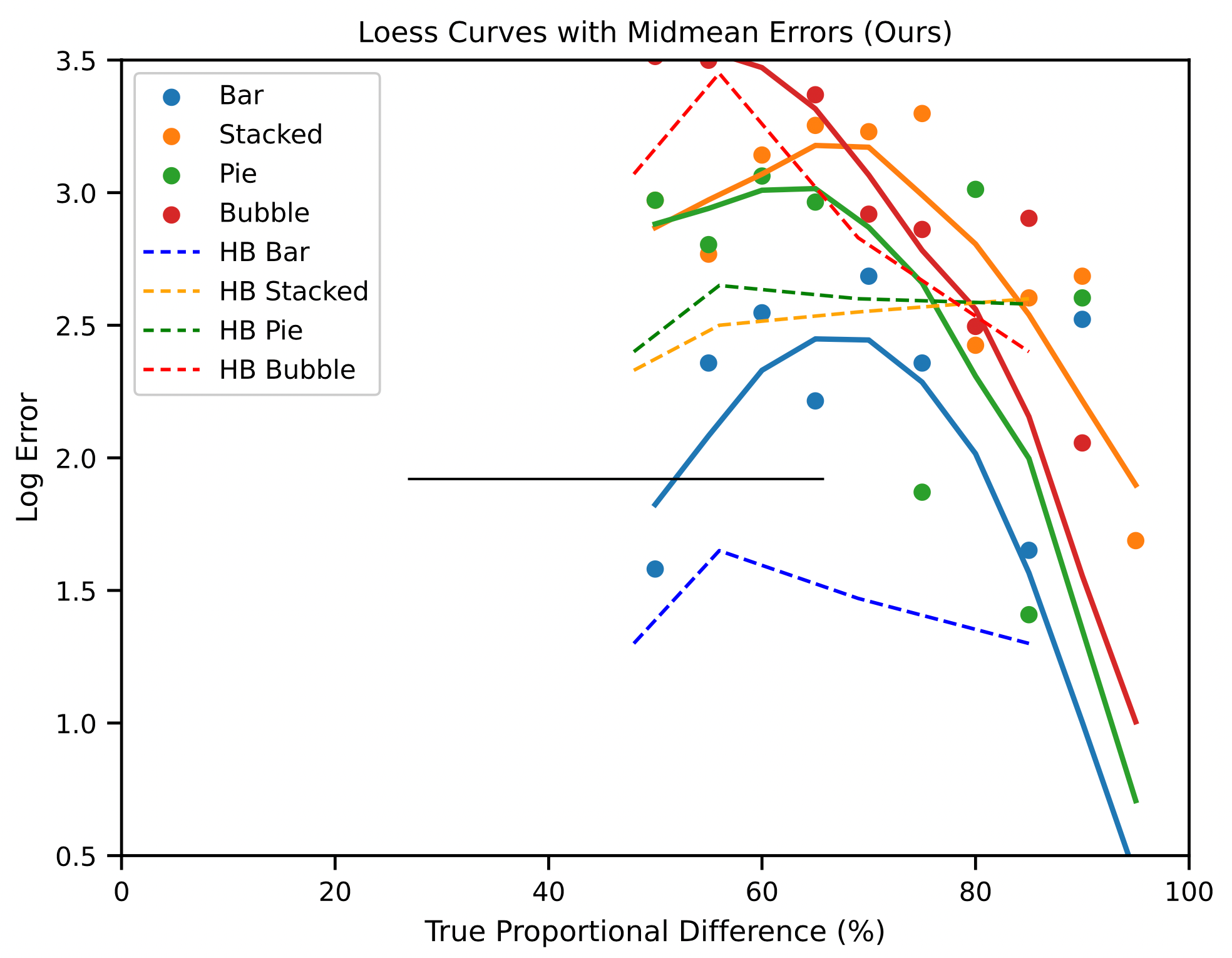}
 \caption{Midmean errors by chart type for ours (solid lines) and Heer and Bostock (2010, sighted participants).  Both datasets show the same general trend, but midmean errors are lower in Heer and Bostock for bar charts and stacked bar charts, a reversal of the overall errors in Figure~\ref{fig:error_rates}, suggesting outliers may have had an effect in our study.  }
 \label{fig:midmeans}
\end{figure}



To facilitate comparison with the original Cleveland And McGill study, we calculated both midmeans for absolute accuracy by ground truth per chart and means and 95\% confidence intervals of error scores for each four bar charts.  Confidence intervals are calculated through bootstrapping.  Error is scored as the log of the absolute difference plus $\frac{1}{8}$, which is added for numerical stability.

\[
\text{Error Score} = \log\left(\left|\text{guessed ratio} - \text{true ratio}\right| + \frac{1}{8}\right)
\]

\noindent
\textbf{Error Score: }
We compare the error scores of participants in our studies to two similar studies: the original Cleveland and McGill study, which was conducted in laboratory conditions in person, and a more recent replication by Heer and Bostock which was conducted remotely and asynchronously over a crowdsourcing platform~\cite{cleveland1984graphical, heer2010crowdsourcing}. 

We note, however, that these comparisons are exploratory and not statistically significant. Our study was not designed to directly replicate results of Cleveland and McGill’s work, but rather to explore the strategies employed by BLV participants.  While we reference their findings to contextualize our results, key differences in participant demographics (e.g., BLV vs. sighted) and task modality (tactile vs. visual) mean that these results should not be interpreted as a one-to-one comparison. In particular, accuracy is dependent on the underlying True Proportional Difference of the data being visualized in the graphical perception experiments.  As a result, our choice of True Proportional Differences, between 50\% and 95\%, could have uniformly deflated our participants' error scores.  However, we still include a rough comparison to previous studies to investigate areas of potential divergence between visual and tactile perception that warrant further investigation.

The directionality of error scores of our study, seen in Fig~\ref{fig:error_rates}, are largely comparable with previous results on sighted individuals found in both these studies.  In particular, length measurements in bar charts had much lower error than the other three explored chart types.  Our results have larger confidence intervals, which is partly explained by a smaller number of trials in our study ($n=651$) vs. the other two  ($n=3481$ for the crowdsourced and $n=2550$ for the laboratory experiment).  However, there are other possible explanations; we analyze the between-subject variance as a source of uncertainty in the subsequent section.

\noindent
\textbf{Midmeans:}
The midmeans chart (Fig~\ref{fig:midmeans}) displays the middle quartiles of error for each group of ground truth, and was used in previous studies to communicate performance without being influenced by outliers.  It indicates that the bar chart is the most accurate chart in general.  Across all chart types, the error decreases as the true proportional difference gets close to 100\%, with a likely peak in error near 50\%. Our results provide evidence that, from most to least effective, length with a common axis, length without a common axis, angle, and area were the most accurate tactile encodings—the same order found in Cleveland and McGill’s original study of visual perception. This matches previous findings reported in a crowdsourced study with sighted participants~\cite{heer2010crowdsourcing}.  However, while the ranking of different charts is similar, there are some differences: the performance on the stacked bar and bar charts had higher midmean error in our study.  This seemingly conflicts with the mean error in aggregate seen in Fig~\ref{fig:error_rates}.  However, the lower error demonstrated in that figure is likely the result of the range of True Proportional Differences in our study limited to 50\% to 95\%, a region which has demonstrated lower error in previous studies.  

\noindent
\textbf{Trial Time:}
Lastly, we also report the average time per trial for our participants and compare to previously reported times.
We compare our results from those reported in Heer and Bostock.  Our participants, on average, viewed 59 charts making the total number of readings $n=651$. The average completion time per chart judgment was 26.74 seconds for BLV users interacting with tactile graphics. In contrast, Heer and Bostock's MTurk study reported a higher average time of 54 seconds per trial, with a median response time of 42 seconds and a standard deviation of 41 seconds.  This comparison suggests that our tactile graphics, while designed for BLV users, enabled relatively faster responses compared to the MTurk study.  We note that Heer \& Bostock suggest that in a laboratory setting, compared with MTurk, they expect that participants would be faster.

\subsection{Hierarchical Analysis}


A work by Davis et al. offers a different type of analysis of graphical perception studies, enabling statistical statements about the variance between subjects using hierarchical modeling~\cite{davis2022risks}.  We repeat that analysis with our results to better understand the outcome of our experiment.

Davis et al. build a hierarchical model by starting with a simple linear model, which assumes that the average performance of each participant is normally distributed about the true mean for the population - essentially that any differences between participants are the result of random chance. They then expand their model by changing different assumptions, which include restrictions on the output of the model (replacing the normal distribution with the Zero-Inflated beta distribution) and adding an additional \textit{random effects} term that assumes some individual differences effect between participant and visualization.  

We reproduce their analysis on our data, and then compare our analysis to their study.  We note that the number of participants (eleven vs. more than a hundred) and the number of trials per participant (approximately sixty vs. more than a thousand) results in experimental datasets of different scales, and so some of the analysis of the resulting hierarchical models still had high levels of uncertainty. But we highlight two charts that can tell us more about the difference between the BLV subjects of our study and the sighted participants in the previous study.

\begin{figure}[h]
 \centering 
 \includegraphics[width=0.5\textwidth]{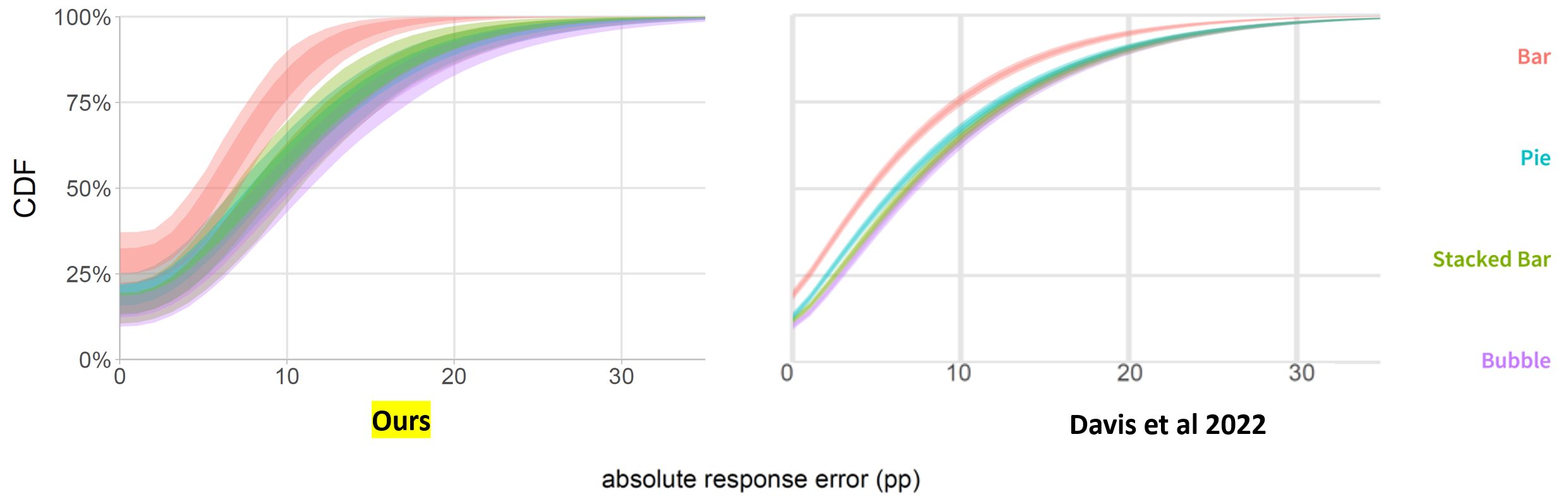}
 \caption{Cumulative Distribution Functions (CDFs) of the hierarchical model comparing our data (left) to data from Davis et al (right)~\cite{davis2022risks}.  
 The difference in shapes suggests that our study resulted in fewer small error estimations, indicated by the flat left side of the curves in our plot.
 }
 \label{fig:cdf_ours}
\end{figure}

First, in Fig~\ref{fig:cdf_ours}, we compare the Cumulative Distribution Functions of the hierarchical models of the absolute response error of the participants' answers.  We are able to produce this chart with the hierarchical model from Davis et al. because it calculates a posterior distribution rather than point estimates, letting us look at the shape of the types of error we found in our study.  It is notable that the both studies have both the same ranking of chart types consistently through different levels of error, both low (less than 5 absolute error) and high (more than 20 absolute error).  At the same time, our study appears to have less errors with very small error (less than 5 absolute error), indicated by the flat region at the very left of the plot.  This is consistent with repeated think aloud statements by our subjects that it was very hard to decide on small differences of estimated answers, i.e. 70\% vs. 75\%.  It is possible that this indicates that the \textit{just noticeable differences} for tactile perception may be larger than visual perception for equivalent graphics, which warrants further study.

\begin{figure}[h]
 \centering 
 \includegraphics[width=0.45\textwidth]{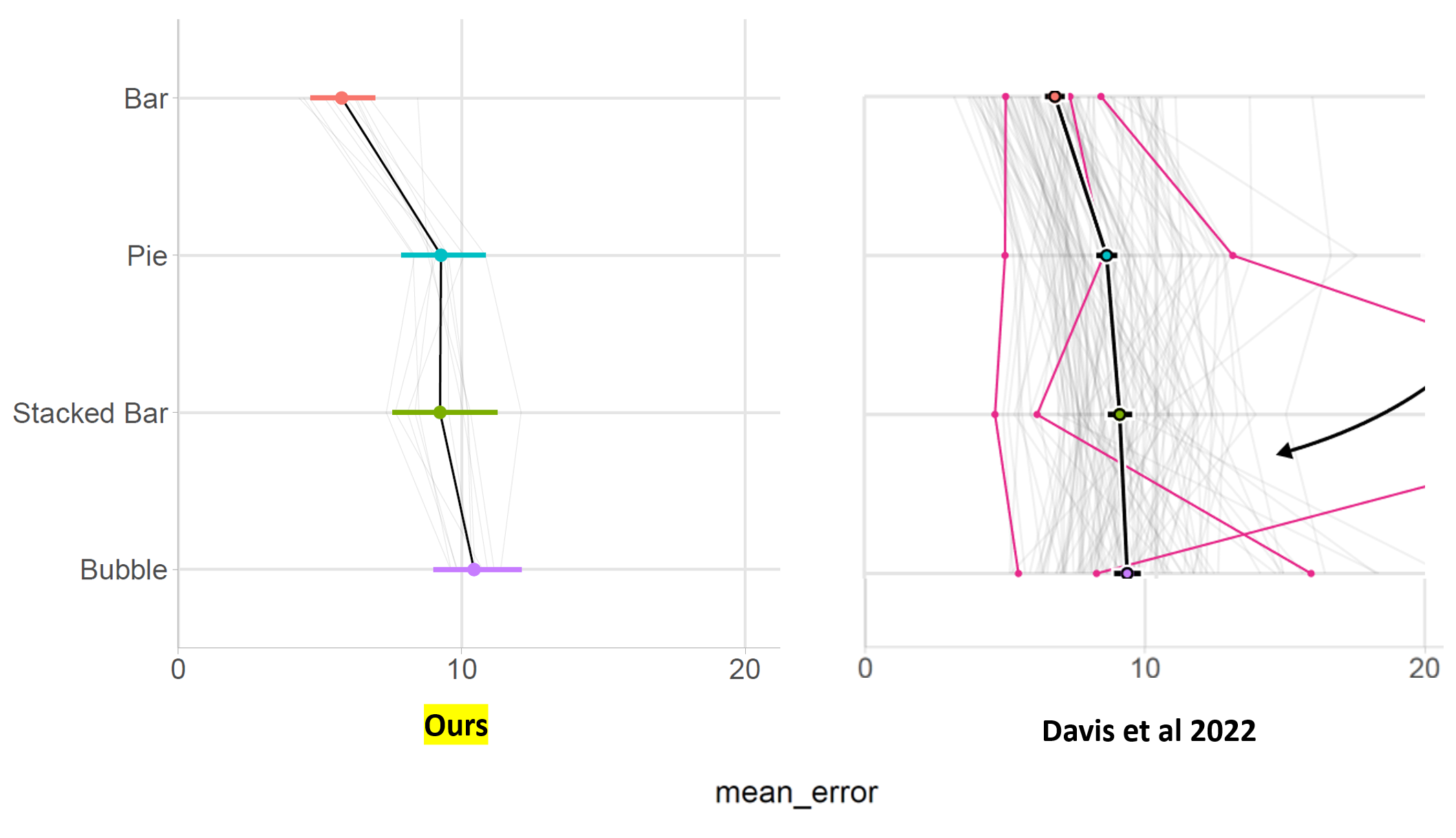}
 \caption{A parallel coordinates chart showing the mean absolute error of our eleven participants, based on a chart from Davis et al~\cite{davis2022risks}.   Results indicate that the performance varies greatly between the most accurate and least accurate participants within our study.  
 }
 \label{fig:parallel_coords}
\end{figure}

Next, in Fig~\ref{fig:parallel_coords}, we compare the mean absolute error across our eleven participants on each chart type with errors from the Davis et al. experiment.  In the parallel coordinates chart, the mean performance of each participant across all 10-20 trials from each of the four chart types is calculated.  This chart was used in Davis et al. to show that some participants have different rankings or relative performances across visualization types - one pink line, for example, showed a bar chart being less effective than a stacked bar chart.  In our data, on the left, we do see similar variance in the rankings of the four chart types within our eleven participants.  It is also notable that the performance appeared to vary broadly in our sample of eleven participants.  However, it isn't clear that the variance noticed in our study was markedly different than the variance evident in Davis et al. Both findings suggest that a larger study would be needed to analyze the types of between-subject and within-subject variance that appears in the data, and to compare it with statistical significance to previous studies.

\subsection{Summary and Takeaways}
Our statistical models reveal trends similar to prior visual perception work—such as higher accuracy for length-based encodings and greater error for area and angle—but these should be seen as hypothesis-generating due to our small, diverse sample. Still, our error analysis suggests Cleveland and McGill’s ordering may hold in tactile contexts, albeit with more variability. Hierarchical modeling showed that length (with common axis) consistently led to lower errors, indicating that future tactile designs might prioritize these encodings. Larger studies with more varied tasks are needed to confirm these findings.

\section{Qualitative Analysis of Subject Strategies}

Through think-aloud statements during the experiment, ad-hoc interviews after the experiment with all eleven participants, and the followup group interview with four participants, we gathered qualitative data about the the strategies they employed as well as difficulties encountered by our subjects that led to error in perceptual judgment.  



\subsection{Strategies}  

A \textit{strategy} was an observed physical interaction between at least one participant and the tactile graphics.  Strategies were sometimes described during thinkaloud, and all strategies were discussed with participants during the end of individual interviews and during the group discussion.  

During think aloud, participants used the terms \textit{calipers} to refer to the measurement of lengths using the space between thumb and forefinger and \textit{ruler} to refer to the measurement of lengths using the number of finger widths to represent a measurement.  Three additional strategies beyond these two measurements of lengths were discussed by our participants, and we extend the metaphor of physical measurement tools to provide evocative names for the strategies, described below.  This results in five identified strategies of tactile perception utilized within our four chart types: \textbf{caliper} and \textbf{ruler} to measure length, \textbf{bevel gauge} to compare angles, \textbf{paint can} to measure area, and \textbf{stud finder} to identify highlighted data items.



These strategies can be broadly categorized into two groups: those that leverage clear measurement analogs to physical tools and those that rely on perceptual heuristics without a direct measurement equivalent. Below, we describe each strategy, mapping them to their closest physical metaphor and discussing their implications for tactile graphical perception.


\begin{figure*}[ht!]
    \centering
    \includegraphics[width=0.8\linewidth]{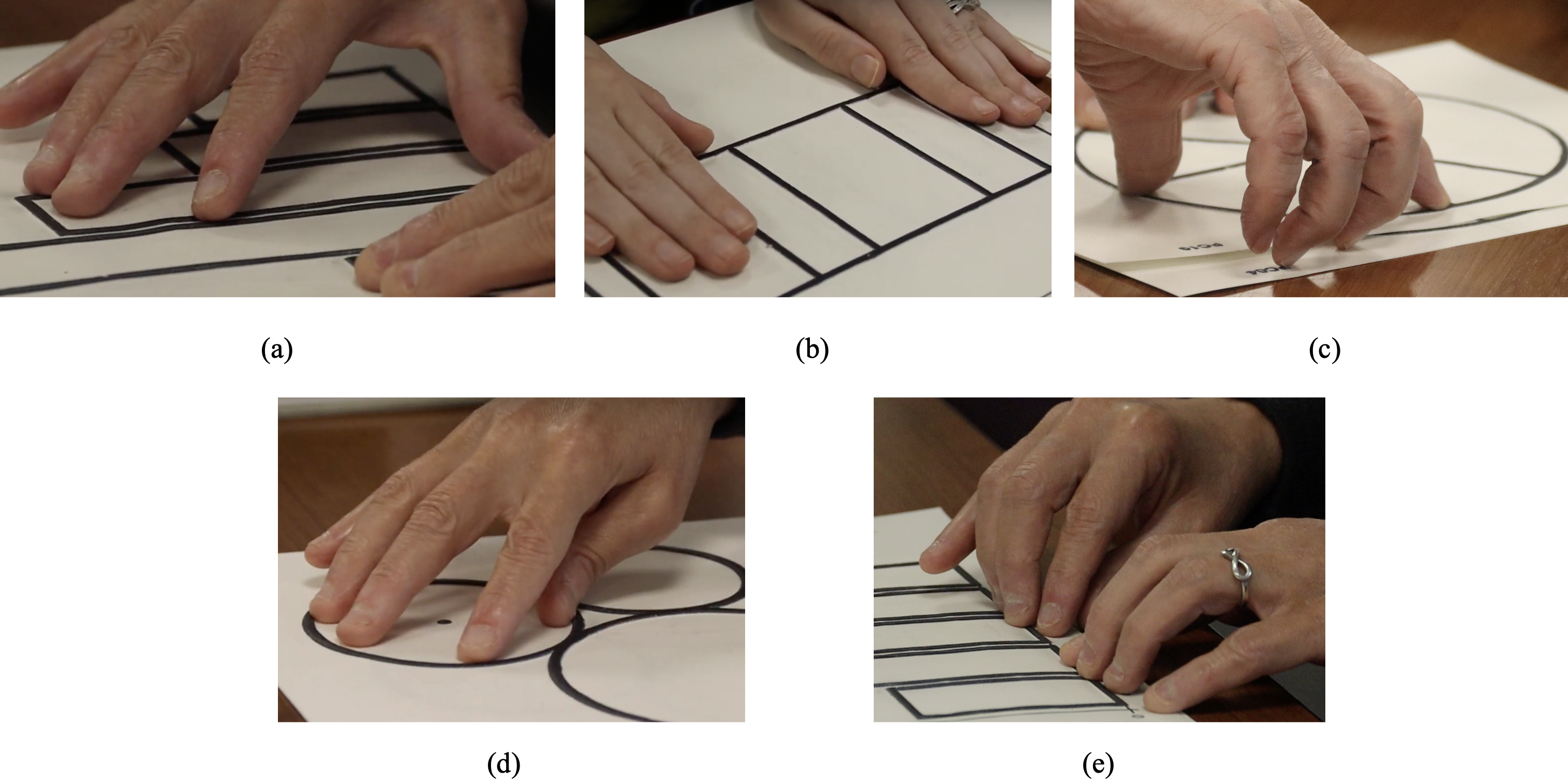} 
    \caption{Pictures of participants using the five identified strategies for tactile perception on our four chart types.  Three strategies ((a) Calipers, (b) Ruler, and (c) Bevel Gauge) are able to directly measure the encoding, while (d) Paint Can and (e) Stud Finder rely on implicit measurements, indicating improvements can be made to tactile design.}
    \label{fig:caliper_strategy}
    \label{fig:finger_measure_strategy}
    \label{fig:studfinder_strategy}
    \label{fig:splaying_finger_method}
    \label{fig:bevelgauge_strategy}
\end{figure*}

\subsubsection{Strategies with Clear Measurement Analogues}
Some strategies directly parallel established physical tools, allowing for precise tactile measurements similar to their visual counterparts. These strategies primarily involve measuring distances, lengths, and angles:

\begin{wrapfigure}{L}{0.8cm}
    \vspace{-\baselineskip}
\includegraphics[width=0.8cm]{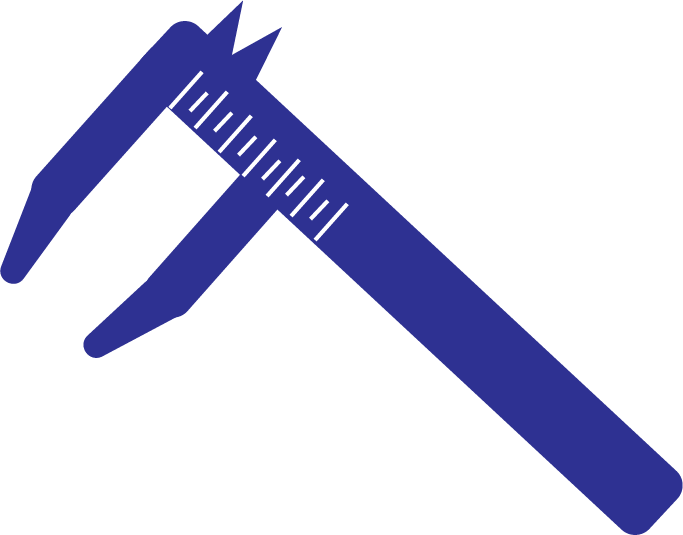}
\end{wrapfigure}

\paragraph{Calipers (Measuring Distance and Proportions)} Participants frequently used their index finger and thumb as a makeshift caliper to measure distances or proportions between graphical elements. In traditional use, a caliper is a fixed tool used to measure the distance between two opposing sides of an object. Similarly, participants would fix the span between their thumb and index finger to match the length of a graphical element—such as the height of a bar or the diameter of a circle—and then “swing” or move their hand to another element to make a comparative judgment. This movement required maintaining a consistent hand shape and orientation throughout the comparison, which posed some challenges.  Figure~\ref{fig:caliper_strategy}(a)  shows a participant using this strategy during the tactile chart reading task. 
This technique was commonly employed in bar charts, stacked bar charts, and bubble charts to assess relative sizes. One participant described:


\say{\textit{I used my fingers kind of across from edge to edge and then tried to move my hand. I usually start with a smaller one and then bring it down to see how much of my finger span filled the bigger one.}}

While this approach aligns closely with visual strategies for estimating proportion and length, it introduces tactile-specific constraints. Participants were limited by the physical span of their fingers, making it difficult to compare elements with large size differences. Additionally, maintaining the same hand orientation during the transition from one element to another was prone to error, particularly in non-linear chart layouts like bubble charts.

\begin{wrapfigure}{L}{0.8cm}
    \vspace{-\baselineskip}
\includegraphics[width=0.8cm]{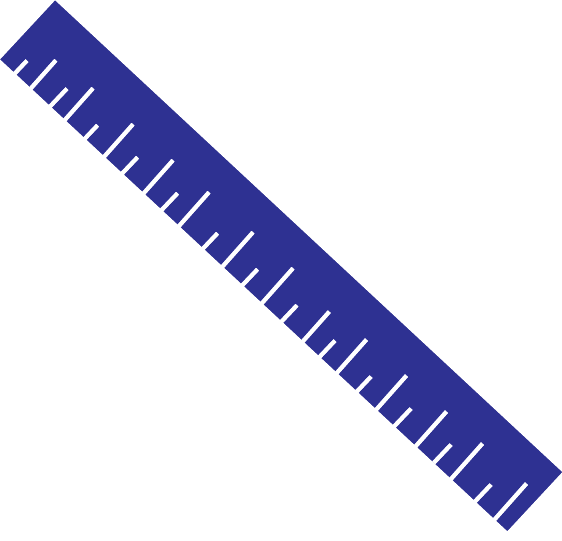}
\end{wrapfigure}
\paragraph{Ruler (Measuring Length Using Fingers as Markers)} Another common approach involved treating fingers as fixed reference points, akin to a tactile ruler, as seen in Figure~\ref{fig:finger_measure_strategy}(b). Unlike calipers, which are used to compare distances directly between two elements, rulers are typically used for counting units along a scale. Similarly, participants in our study used individual fingers—or segments of their fingers—as unit markers to estimate the height, width, or length of visual elements.

This technique was particularly useful for measuring linear dimensions in bar charts and stacked bar charts, where the shape aligned well with the span of the fingers. 
One participant explained:


\say{\textit{I started using my fingers as my measuring tools for bar charts.}}

While this method helped participants make consistent judgments, it also introduced limitations. Finger sizes vary between individuals, and there are only a finite number of usable markers on one hand. Participants often encountered difficulties when values fell between finger widths, requiring mental interpolation or the use of "half fingers" to estimate values. As one participant noted during the group interview:


\say{\textit{There's no reason why this couldn't be a ruler that you count up 1, 2, 3, 4—and a half—high.}}

This illustrates how participants creatively adapted their own body as a measurement tool, while also highlighting the imprecision inherent in this strategy.

\begin{wrapfigure}{L}{0.8cm}
    \vspace{-\baselineskip}
\includegraphics[width=0.8cm]{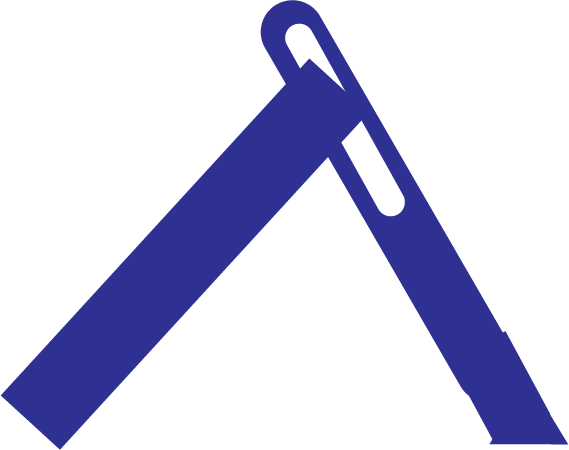}
\end{wrapfigure}

\paragraph{Bevel Gauge (Measuring Angles and Orientation)} When interpreting pie chart wedges, participants used their fingers to approximate angles, effectively replicating the function of a bevel gauge. A bevel gauge is a measuring tool with an adjustable blade that can be fixed at a desired angle relative to a base, allowing users to replicate or compare angular measurements. Similarly, participants placed their thumbs and index fingers along the inner edges of a wedge, spreading their hand to estimate the angular size through proprioception. 
One participant noted:


\say{\textit{When I found the dots, I instantly went into the wedges, and kind of went to the tips where they meet, then traced my thumbs and index fingers along the inner lines of the wedges to get a proprioceptive value of how far I was spreading one hand to the other.}}

While this method supported rough comparisons between wedges, it had notable limitations. Participants often avoided using the full span between thumb and forefinger, finding it too imprecise for subtle distinctions. Instead, many relied on the tips of their fingers to trace along the edges, which provided more control but covered less distance, reducing the accuracy of the comparison. This strategy, though intuitive, was sensitive to slight shifts in hand orientation and lacked consistent reference points for repeated measurements.
Figure~\ref{fig:bevelgauge_strategy}(c) shows a participant employing this technique during a pie chart interpretation task.



\subsubsection{Strategies Without Direct Measurement Analogues}

Other strategies emerged in cases where no clear physical tool existed to support measurement, requiring users to rely on estimation and heuristic-based approaches. These strategies highlight the perceptual differences between tactile and visual graphical interpretation, and so for each strategy, we describe limitations in existing visual encodings that are measured by them.

\begin{wrapfigure}{L}{0.5cm}
    \vspace{-\baselineskip}
\includegraphics[width=0.5cm]{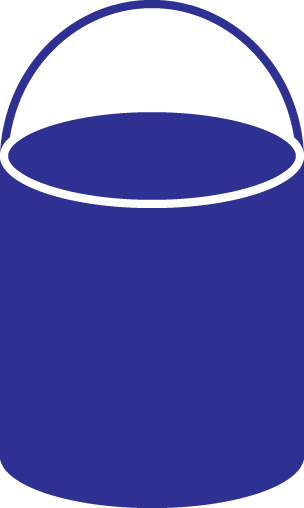}
\end{wrapfigure}


\paragraph{Paint Can (Estimating Area Using Splayed Fingers)} When evaluating area-based encodings, such as bubble charts and pie chart wedges, participants adopted a strategy reminiscent of estimating paint coverage with a brush. Like gauging how much paint is needed to cover a surface, this method involves an indirect sense of volume or spread rather than precise measurement. Participants splayed their fingers outward and swept them across the surface of the shape to develop a general sense of its spatial footprint. 
Figure~\ref{fig:splaying_finger_method}(d) shows a participant using the paint can strategy, sweeping their splayed fingers across the shape to estimate its area through tactile exploration.
\say{\textit{I tried spacing my finger on the edges of each of the wedges and tried to see how that space could be divided into smaller sections or see how many fingers would fit in each.}}

\noindent
\textbf{Limitations:} 
This strategy, while helpful for building a mental impression of relative size, required participants to mentally convert their physical gestures into estimations of area—a process that relies heavily on short-term memory. Because the entire shape could not be taken in at once, participants often needed to retain impressions from prior exploration and compare them mentally, rather than through direct measurement. Although some attempted simultaneous comparison using both hands, this was rare and difficult to execute with accuracy. Compared to strategies like the bevel gauge, the “paint can” method was considerably less precise, particularly when making fine-grained comparisons between similarly sized regions. 

\begin{wrapfigure}{L}{0.35cm}
    \vspace{-\baselineskip}
\includegraphics[width=0.35cm]{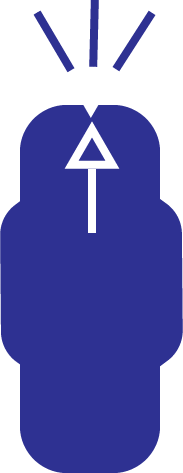}
\end{wrapfigure}
\paragraph{Stud Finder (Locating Target Graphical Primitives)} Identifying the relevant data points in tactile graphics often required a broad exploratory sweep before engaging in detailed measurement. A stud finder is a handheld device used to detect studs behind a wall by scanning across a surface and signaling when material changes are detected. Similarly, participants used their fingertips to scan across the tactile surface, identifying areas that stood out texturally or spatially before zeroing in for closer analysis as seen in Figure ~\ref{fig:studfinder_strategy}(e). One participant described this approach:

\say{\textit{I generally first would look for the 2 dots to identify the bars that I needed to look at and then examine both bars.}}

\noindent
\textbf{Limitations: }
This behavior indicates that BLV users require distinct affordances to efficiently locate key graphical elements, particularly when interacting with complex visualizations.  This can be an opportunity for improved design, as popout effects for BLV users may be different than sighted users.

\subsubsection{Difficulties}

Participants described their frustrations throughout think aloud, and were explicitly asked about what made it difficult to read the  charts.  We broadly categorize these difficulties into challenges with \textbf{scale}, \textbf{distance}, \textbf{tactile noise}, \textbf{local vs. global} sense of the data, and \textbf{visual memory}.   

\noindent
\textbf{Scale:} For scale, we note that both the calipers strategy and the ruler strategy have limitations in the scale of graphics they can measure.  While visualizations for sighted viewers have limitations of scale as well, reasonable scales may be much larger (i.e. large displays) and much smaller (i.e. Apple Watch), than the reasonable scales for tactile graphics.  Participants mentioned that having smaller hands felt like a disadvantage for some of the larger graphical primitives being read.  

\noindent
\textbf{Distance: }The amount and orientation of the distance between marks being measured also added additional difficulties, particularly with the caliper, bevel gauge, and paint can strategies, which often required the measuring hand to move from one mark to the other to make a comparison.  The longer this distance was, the greater chance of accidentally shifting the measurement of the caliper.  This was sometimes exacerbated by marks that had orientations that were not axis-aligned, as in the bubble chart and pie chart.

\noindent
\textbf{Tactile Noise: }Starting from Tufte's rules about maximizing the data-ink ratio~\cite{tufte1983visual}, it is accepted that additional noise in the visual encoding can distract from the underlying perceptual task.  Our participants mentioned that the design of our graphics contained noise that made it difficult to focus on or even locate the particular marks being measured.  However, the types of noise that were identified were not anticipated.  First, participants noted that the width of the lines in our swell form graphics were disruptive, and that they would prefer thinner lines.  It was also noted that the white space between marks was possibly misleading, because the positive space inside the marks and the negative space between marks may be harder to discriminate in tactile perception.  The bar chart was particularly challenged by white space, since the space between marks is fully redundant.  Lastly, because participants can only use part of their hand at once and do not interact with the entire marks, the marks themselves may not have needed their full tactile encoding.  This was most prevalent in the bubble chart, where the areas where the circles touch are the most noisy areas, but those areas were not typically used by participants.  It may be that space-filling visualizations like bubble charts are not effective for tactile visualizations because of the noise resulting from the cramped and unaligned layout of marks.  In general, the heuristics and gestalt rules~\cite{wong2010points} taken for granted in visual design should be interrogated for tactile design.

\noindent
\textbf{Local vs. Global: }A broader difficulty that was experienced by our participants was the relationship between a \textit{global} view of the data and a \textit{local} view of the data.  Shneiderman's mantra (\textit{overview first, zoom and filter, details on demand}), a motivating heuristic for visualization design, states that a global view of the data is usually the first impression that a user wants~\cite{shneiderman2003eyes}.  However, one participant brought up independently that they did not typically generate a global perception of the data. 


\say{\textit{When you're using your fingers to do things, your hand is covering things you can't be viewing.  We can't look at these things globally.  We have to look at them in a micro- rather than a macro- kind of way.}}

This suggests that elementary perceptual tasks like those used in this study and Cleveland and McGill, which are typically local, might be expected to be comparable between sighted and visually limited participants, but more global tasks could result in a wider gap and greater need for new encodings for tactile graphics.  The quote also speaks to potential \textit{Midas Touch} issues~\cite{velichkovsky1997towards}, where the interaction with a graphic may intersect with the ability to use that graphic for input, in a theoretical interactive tactile system.

This suggests that global perception through touch may require different cognitive and perceptual strategies than visual overview. Gestalt properties that support transitioning from global to local analysis in visual graphics may not translate directly to tactile formats. This distinction has implications for the design of interactive systems, particularly those that rely on global-first paradigms like Shneiderman’s mantra. Prior work (e.g., Watanabe and Mizukami \cite{WatanabeMizukami2018}; Perdue et al. \cite{Yalong2020}) shows that users can and do engage in global pattern recognition with tactile charts—especially when relationships or trends (e.g., in scatter plots) are central to the task. 
Therefore, our observations do not imply that tactile graphics are inherently poor for global tasks, but rather that additional design considerations and affordances may be needed to support them.

\noindent
\textbf{Visual Memory: }Lastly, participants frequently brought up the significance of having visual memory for perceiving graphical primitives.  They described that it is generally well known within the BLV community that those who become blind during their lives often have visual memories that can improve their ability to interpret visual concepts.  One participant of our group interview was born blind, rather than becoming blind during their life, and remarked at how they found the process of reading tactile graphics very frustrating.


\say{\textit{I've never seen a graph, I've never seen anything visually.  I may be frank -- graphics mean nothing to me.  I have no context, so I would dismiss them out of hand if \textbf{they aren't built for me}...  If they're built from a visual place, at least for me, they mean nothing, they have no resonance with me whatsoever.}}

This suggests that additional studies are necessary to understand whether the presence of visual memory impacts the perceptual accuracy of BLV subjects.  It also suggests that a user-driven design of tactile graphical primitives designed primarily for those born blind could result in more effective encodings for tactile graphics.









\subsection{Gaps}
\label{sec:gaps}


While the strategies employed by BLV users enabled effective interpretation of tactile graphics, they also revealed significant limitations in current visualization affordances. We synthesize our empirical results, our analysis of participant strategies, and the difficulties reported by participants, along with existing visualization research to identify gaps between the needs for users and the design guidelines for visualizations.  These gaps highlight both the physical constraints imposed by tactile interaction and the cognitive load required for interpretation, underscoring opportunities for more accessible design interventions.


\noindent\textbf{G1: Physical Constraints on the Size and Orientation of Graphical Primitives}
One of the most fundamental limitations observed was the constraint imposed by hand size on tactile measurement strategies. Unlike visual perception, which allows for rapid scanning and simultaneous comparison of multiple elements, tactile interaction is constrained by the spatial extent of the hands. This limitation manifests in several ways:

\begin{itemize}
    \item \textbf{Calipers Limited by Hand Size:} Participants using their thumb and forefinger as calipers faced a natural upper bound on the range of distances they could measure. This was particularly challenging when comparing larger graphical primitives.
    
    \item \textbf{Ruler Length Bound Above by Hand Size, and Precision Bound Below by Finger Size:} Participants who used their fingers as rulers faced an inherent limitation where the length of their fingers restricted their ability to compare larger objects, while the width of their fingers imposed a lower bound on precision.

    \item \textbf{Bevel Gauge Constraints in Measuring Angles:} When attempting to estimate angles, participants preferred to use their fingertips rather than the space between their thumb and forefinger, as the latter was not sufficiently precise. 

    \item \textbf{Two-Item Comparison Limitation:} Participants were limited to comparing only two items at a time due to the use of both hands. Unlike sighted users who can rapidly compare multiple elements in parallel, BLV participants had to engage in sequential comparisons.

    \item \textbf{Differences in Distance Traversal:} In visual perception, the eyes can rapidly scan across distances without physical effort, whereas in tactile perception, participants must physically trace elements with their hands.

    \item \textbf{Rotations and translations are necessary for measurements:} For stacked bar charts and pie charts, participants frequently rotated the chart or moved the sheet completely to make the physical measurements more comfortable from their seating position.  An example can be seen in the participant's reading of the stacked bar chart in Fig~\ref{fig:finger_measure_strategy}, where the fingers pointing away from the body should be perpendicular to the axis being measured.  The orientation of the graphical primitives appears to be important for tactile perception in a unique way compared to visual graphical primitives, and should be analyzed in a future study.

\end{itemize}

\noindent\textbf{G2: Absence of a Clear Measurement Tool for Area Estimation}

While participants had well-defined strategies for measuring distances and angles, no equivalent strategy existed for estimating areas. This gap resulted in significant cognitive load, as participants had to rely on memory and approximation to derive area comparisons.

\begin{itemize}
    \item \textbf{High Cognitive Load for Tactile Area Estimation:} Measuring areas often required either breaking the shape down into measurable lengths and mentally calculating the area or using the "paint can" strategy—sweeping their fingers over a region to gauge coverage. This approach introduced substantial cognitive strain. One participant expressed this difficulty:
    \say{\textit{
        If you measure the diameter of a circle, I wouldn’t know how to calculate the area. I don’t know the formula.}}

\end{itemize}

\noindent\textbf{G3: Localized Exploration Hinders Object Identification}

Unlike visual users who can maintain a simultaneous global and local perspective on a chart, BLV users engage in tactile exploration in a localized manner. This makes it difficult to establish spatial relationships between elements, leading to challenges in quickly identifying key objects.

\begin{itemize}
    \item \textbf{Difficulties in Maintaining a Global Perspective:} Participants were frustrated by only being able to examine a small section at a time. One participant described:
    \say{\textit{When you’re using your fingers, your hand is covering things you can’t be viewing. We can’t look at these things globally. We have to look at them in a micro rather than a macro kind of way.}}

    \item \textbf{Spatial Distance Affects Object Recognition:} Because data exploration was localized, identifying relationships between elements positioned far apart was challenging. This reinforces the need for texture-based or interactive solutions to bridge these perceptual gaps.

\end{itemize}

We posit that these gaps exist because traditional visualizations are designed to leverage visual strategies such as parallel scanning, rapid perceptual grouping, and spatial memory that do not directly translate to tactile exploration. Unlike visual charts, which allow for holistic views and quick comparisons, tactile graphics require sequential exploration, increasing cognitive load and limiting simultaneous access to multiple data points. Furthermore, visualization tools often lack built-in affordances that support tactile estimation of measurements such as area, distance, or relative positioning.

\subsection{Guidelines for Tactile Graphics Design}
\label{sec:guidelines}

Our findings suggest a set of design guidelines for improving the accessibility and effectiveness of tactile graphics for BLV users. We divide these guidelines into two categories: (1) those with known solutions that can be implemented based on existing research, and (2) open problems in tactile representation design that remain unsolved.

\subsubsection{Guidelines with Known Solutions}

Several strategies have been successfully used in tactile and visual graphics that can be adapted to improve tactile data representation.

\noindent
\textbf{Direct Encoding of Area Values: }
One of the primary challenges faced by participants was accurately comparing areas. To mitigate this issue, we recommend encoding numerical values directly within area-based graphical elements. This can be achieved through secondary encoding techniques such as embedding braille text inside shapes. This approach is inspired by dashboard design, where numerical values are often placed within bar charts for emphasis. By providing both graphical representation and encoded values, users can access details on demand while preserving the graphical structure.

\noindent
\textbf{Double Encoding for Improved Saliency: }
Tactile encodings often struggle with visual pop-out effects, making it difficult to distinguish elements. Double encoding—combining textures with raised lines, varying line thickness, or using filled vs. unfilled areas—can improve perceptual saliency. 


\say{\textit{Use a filled in texture as opposed to the smooth texture [for the dotted parts]. I think it would be easier to compare and contrast them, than just having the openness of the two lines and the rectangles, if they were filled in, it would be spot on.}}

Using texture differentiation and redundant encodings can help users locate and compare graphical elements more efficiently.  Appropriate use of textures for fill patterns were proposed in a previous study by Engel and Weber~\cite{Engel2017}.

\noindent
\textbf{Pixel-Aware and Spatial-Aware Encodings: }
Swell form printing has physical limitations on the size and weight of strokes, and newer media like refreshable tactile displays (RTDs) typically have low resolution, requiring thick lines for detectability. This constraint often leads to cluttered representations. Prior research on pixel-aware encodings suggests that encoding density and resolution-aware techniques must be adapted for tactile graphics~\cite{szafir2017modeling}.

Additionally, spatial constraints impact usability, as graphical elements must fit within a user’s hand span to be effectively compared. Future tactile encodings should optimize spacing and resolution to enhance clarity while minimizing tactile noise.

\noindent
\textbf{Local, Not Global: Minimizing Distractors and Reducing Distance Between Comparisons: }
BLV users rely on localized tactile exploration rather than global scanning, making it crucial to remove distractors and place key comparisons in close proximity. Hierarchical visualizations that allow users to drill down for details may be more effective than complex static layouts
This suggests a need for hierarchical views with collapsible elements, allowing users to focus on relevant data while still accessing broader context when needed.

\subsubsection{Guidelines with No Known Solutions}

Several tactile visualization challenges remain unresolved. These open problems highlight the need for further research and innovation.

\noindent
\textbf{Interactivity in Low-Resolution Refreshable Tactile Displays: }
Many participant challenges stemmed from the limited resolution of RTDs, which restricts detailed representations and interactive feedback. While interactive techniques have been explored in visual analytics, their adaptation to tactile graphics remains an open problem.  Familiar paradigms in visual interactive systems typically rely on high resolution large screens~\cite{fink2009visualizing} to present multiple views of the data featuring coordinated multiple views~\cite{roberts2007state}.  It remains unclear how to build such systems with low-resolution dynamic displays in tactile formats.  Future research should investigate haptic interactions that provide additional information without overwhelming the user’s sense of touch.

\noindent
\textbf{Addressing the “Midas Touch” Effect: }
A common issue in tactile graphics is the occlusion of key elements by the user’s own hands. This is analogous to the "Midas Touch" effect in eye-tracking research, where interaction is unintentionally triggered by gaze fixation. Developing techniques that allow for non-intrusive exploration remains a challenge.  One possible solution could involve multi-stage feedback that distinguishes exploratory from selection-based interactions.

\noindent
\textbf{Multiple Views and Details on Demand: }
Visual analysis tools allow users to access multiple perspectives on the same data~\cite{chen2020composition}, yet tactile equivalents remain limited. Introducing multiple views, such as hierarchical summaries or separate secondary displays for detailed information, could improve usability.  Concepts such as "draggable snapshots" or comparison drawers could facilitate side-by-side evaluation of different data points over time.

\section{Discussion and Future Work}


\subsection{Future of Tactile Graphics}
Our findings suggest that while tactile graphics adapted from visual design principles can be effective, they may not fully meet the accessibility needs of those born without sight. This aligns with recent studies emphasizing the importance of designing with and for BLV users to create truly effective and accessible tactile graphics \cite{zong2022rich, lundgard2021accessible}. As \cite{reinders2024refreshable} suggests, incorporating conversational agents and advanced tactile displays could further enhance accessibility, pointing towards a multimodal approach as a promising future direction. Our results support this direction, indicating that tactile graphics should be part of a broader, inclusive design strategy prioritizing user-centric development.

\subsection{Limitations}
While it produced useful insights, our study had some limitations. The primary constraint is the reliance on existing visual encodings which might not optimally translate into tactile formats. This limitation is evident in our mixed results where some tactile graphics performed well while others did not, indicating a possible mismatch in the tactile adaptation of visual data \cite{kim2023exploring}. Additionally, our sample size and the diversity of our participant group, though adequate for preliminary insights, may not fully represent the wider BLV population, possibly limiting the generalizability of our findings.

The design of tactile graphics itself, produced using swell paper technology, introduces another limitation due to the restricted resolution and detail this method can provide. This potentially affects the users' ability to discern fine details in complex graphs, potentially skewing the accuracy and speed of data interpretation when compared to visual graphs as noted in foundational studies \cite{cleveland1984graphical}. The accessibility and cost of producing high-quality tactile graphics also pose significant limitations. While swell form technology is cost-effective, it is not universally accessible, and the financial burden may be prohibitive for some institutions or individuals, limiting widespread adoption.

Moreover, the diversity of techniques used by participants to interpret the data underscores a limitation in the standardization of tactile graphics. For example, some participants primarily used the caliper strategy for measuring length, while other primarily used the ruler strategy. Such personalized techniques, although creative, highlight the lack of a one-size-fits-all approach in the current tactile graphic designs.

Participants also noted difficulties with the texture and construction of tactile elements. Many found the uniformity in the texture of bars confusing and suggested that varying the texture could aid in differentiation. Double encoding, which combines tactile information with other sensory cues like varied textures, was recommended to improve usability and accuracy.

\subsection{Future Studies}
Our work highlighted some potential solutions, but also uncovered some open problems.  To explore solutions to these open problems, we believe collaborative design studies with BLV participants are appropriate~\cite{ebermann2024sight}.  These studies should explore a wider array of tactile and multimodal presentation techniques.  Collaboration with BLV participants should be emphasized to tailor designs more closely to user needs, potentially involving technologies that integrate tactile feedback with auditory and possibly olfactory cues for a richer user experience \cite{zong2022rich}. We anticipate that such inclusive research efforts will be crucial in addressing the broader accessibility challenges faced by the aging population with vision loss \cite{visionserve2022}.

In addition, we also would like to explore broader participant pools for exploring our findings from our research.  A series of large-scale studies with greater variance in experience level with tactile graphics and educational background could further validate and refine the tactile graphic findings.  

\section{Conclusion}
In this research, we replicated the Cleveland and McGill study on graphical perception using tactile graphics to explore their efficacy for BLV individuals. Our findings suggest that while tactile representations derived from visual design principles hold promise, their efficacy is not universally optimal, particularly for those born without sight (e.g., \cite{reinders2024refreshable}). This highlights the need for designs that explicitly consider the unique perceptual requirements of BLV users.

The exploration of tactile graphics in our study suggests that they can serve as effective tools for data representation for the visually impaired when adapted with consideration for tactile perception. However, the mixed results across different chart types indicate that further refinement is necessary to fully leverage these tools. This aligns with recent advances in multimodal data representation which suggest integrating tactile feedback with auditory and possibly olfactory cues to enrich the data interaction experience \cite{lundgard2021accessible,zong2022rich}.


As we look towards the future, it is clear that tactile graphics will continue to evolve, reflecting the advancements in technology and a deeper understanding of accessible design. It is our hope that these efforts will significantly reduce the barriers faced by the visually impaired community, granting them greater access to the burgeoning field of data science and visual analytics.


\bibliographystyle{abbrv-doi-hyperref}

\bibliography{tactile_graphics_updated}

\appendix 

\end{document}